\documentstyle[sprocl]{article}
\input{psfig.sty}

\def\la{\mathrel{\mathpalette\fun <}}
\def\ga{\mathrel{\mathpalette\fun >}}
\def\fun#1#2{\lower3.6pt\vbox{\baselineskip0pt\lineskip.9pt
  \ialign{$\mathsurround=0pt#1\hfil##\hfil$\crcr#2\crcr\sim\crcr}}}

\begin{document}

\title{TOPOLOGICAL DEFECT MODELS OF ULTRA-HIGH ENERGY COSMIC RAYS}

\author{G\"unter Sigl}

\address{Department of Astronomy \& Astrophysics and
Enrico Fermi Institute\\
The University of Chicago, Chicago, IL~~60637-1433\\
and\\
NASA/Fermilab Astrophysics Center\\
Fermi National Accelerator Laboratory, Batavia, IL 60510}

\maketitle\abstracts{\centerline{Abstract}\vspace{0.25truecm}
We give an overview over models in which cosmic rays above
$\sim1\,$EeV ($=10^{18}\,$eV) are produced by the decay of supermassive
``X''particles released from topological defects possibly
created in cosmological phase transitions. We note that,
for an interesting particle physics parameter range, 
these models are still consistent with current data, and discuss
signatures for the topological defect mechanism which can
be tested by the next generation experiments.}

\section{Introduction}
The highest energy cosmic ray (HECR) events observed above
$100\,$EeV~\cite{Bird,AGASA} are difficult to explain within
conventional models involving first order Fermi
acceleration of charged particles at astrophysical
shocks. On the one hand, even the most powerful astrophysical objects
such as radio galaxies and active galactic nuclei (AGN) are
barely able to accelerate charged particles to such
energies.~\cite{Hillas} On the other hand, above
$\simeq70\,$EeV the
range of nucleons is limited by photopion production on the
cosmic microwave background (CMB) to about $30\,$Mpc (which is
known as the GZK-effect~\cite{GZK}), whereas heavy nuclei are
photodisintegrated on an even shorter distance scale.~\cite{Puget}
In addition, for commonly assumed
values of the parameters characterizing the galactic and
extragalactic magnetic fields, protons above
$100\,$EeV are deflected by only a few degrees over these
distances~\cite{SSB} and obvious powerful sources have not been
found in the arrival direction of the observed HECR
events.~\cite{ES}

\begin{figure}[ht]
\psfig{file=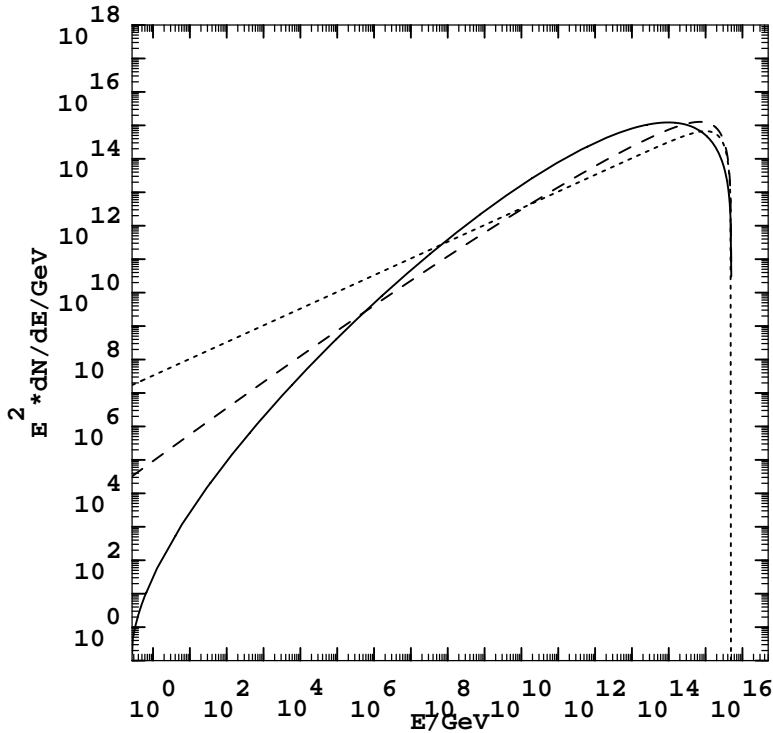,height=3.8in}
\medskip
\caption[...]{Approximations to the hadronic fragmentation function
$dN_h/dE$ for $m_X=2\times10^{16}\,$GeV. The solid line indicates
the solution to the QCD evolution equations in modified leading
logarithmic approximation,~\cite{detal} the dashed line is the
one used in,~\cite{BHS} and the dotted line represents the
approximation $dN_h/dx=(15/16)x^{-1.5}(1-x)^2$ used
in~\cite{Hill,BS,PS} and in Fig.~\ref{F2}. \label{F1}}
\end{figure}

Within ``top-down'' (TD) scenarios, in contrast, predominantly
$\gamma$-rays and neutrinos are initially produced at
ultra-high energies (UHEs) by quantum
mechanical decay of supermassive elementary X particles related
to some grand unified theory (GUT). Such X particles could be
released from topological defect relics of phase
transitions which might have been caused by spontaneous breaking
of GUT symmetries in the early universe.~\cite{BHS}
Topological defects from phase transitions in the early universe such as
cosmic strings, monopoles, and domain walls are topologically
stable, but nevertheless can release part of their energy via collapse
or annihilation in the form of X particles. The X
particles can be gauge bosons, Higgs bosons, superheavy fermions,
etc.~depending on the specific GUT, their mass $m_X$ being
comparable to the symmetry breaking scale. They
subsequently typically decay into a lepton and
a quark which roughly share the initial energy equally. The quark
then hadronizes into nucleons ($N$s) and pions, the latter ones in turn
decaying into $\gamma$-rays, electrons, and neutrinos. 
Given the X particle injection rate, $dn_X/dt$, the effective
injection spectrum of particle species $a$ ($a=\gamma,N,e^\pm,\nu$) can be
written as $(dn_X/dt)(2/m_X)(dN_a/dx)$,
where $x \equiv 2E/m_X$, and $dN_a/dx$ is the relevant effective
fragmentation function. We take the primary lepton to be an
electron injected with energy $m_X/2$.
The total hadronic fragmentation function $dN_h/dx$ can be taken
from solutions of the QCD evolution equations in modified leading
logarithmic approximation which provide good fits to accelerator
data at LEP energies.~\cite{detal} Fig.~\ref{F1} shows this
function for $m_X=2\times10^{16}\,$eV in comparison to approximations
used in earlier work.~\cite{Hill,BHS} Motivated by LEP data at
lower energies,
we assume that about 3\% of the total hadronic content consists of
nucleons and the rest is produced as pions and distributed equally among
the three charge states. The standard pion decay spectra then
determine the injection spectra of $\gamma$-rays, electrons, and
neutrinos. The X particle injection rate is assumed to be
spatially uniform and in the matter-dominated era can be
parametrized~\cite{BHS} as $dn_X/dt\propto t^{-4+p}$,
where $p$ depends on the specific defect scenario.
The case $p=1$ is representative for a
network of ordinary cosmic strings~\cite{BR} and
annihilation of monopole-antimonopole pairs.~\cite{BS}

Since the
absolute flux level predicted by TD models is very model
de\-pen\-dent,~\cite{GK} we will allow an arbitrary flux
normalization noting that certain TD
scenarios such as annihilation of magnetic monopole-antimonopole
pairs~\cite{BS} yield HECR fluxes consistent with observations
without violating bounds on monopole abundances.
Such models are attractive in
explaining HECRs because they predict injection spectra
which are considerably harder than shock acceleration spectra
and, unlike the GZK effect for nucleons, there is no threshold
effect in the attenuation of UHE $\gamma$-rays whose mean free
path in the cosmic low energy photon background is probably
larger than that for nucleons (see, e.g.~\cite{Lee}).

\section{Signatures for and Constraints on TD Models}
Based on the general features of the type of TD scenarios
discussed in the previous section, there are two distinctive signatures for
them at energies above $100\,$EeV: First, the observed
primaries should be predominantly $\gamma$-rays.~\cite{ABS}
Second, there could be a spectral feature in the form of a
``gap''.~\cite{SLSB} Increase of the current total exposure at
these energies by factors of a few could distinguish between
acceleration type sources and TD mechanisms at the 99\%
confidence level.~\cite{SLSB} This should easily be possible
with next generation experiments under construction such as the
High Resolution Fly's Eye~\cite{Bird1} or in the
proposal stage~\cite{Proc} such as the Pierre Auger project.

Recently, there has been considerable discussion in the
literature whether the $\gamma$-ray, nucleon, and neutrino
fluxes predicted by TD scenarios are consistent with
observational data and constraints at any
energy.~\cite{Chi,SJSB,PS}
To thoroughly investigate this, we have performed
extensive numerical simulations for the propagation of
extragalactic nucleons, $\gamma$-rays, and electrons with
energies between $10^8\,$eV and $10^{25}\,$eV through the
universal low energy photon background,~\cite{Lee,SLC1} which
includes the radio background, the CMB, and the infrared/optical
(IR/O) background.
All relevant interactions have been taken into
account, including synchrotron loss in the EGMF of the
electronic component of
the electromagnetic cascades which result from UHE $\gamma$-ray
injection into the universal radiation background.

\begin{figure}[ht]
\psfig{file=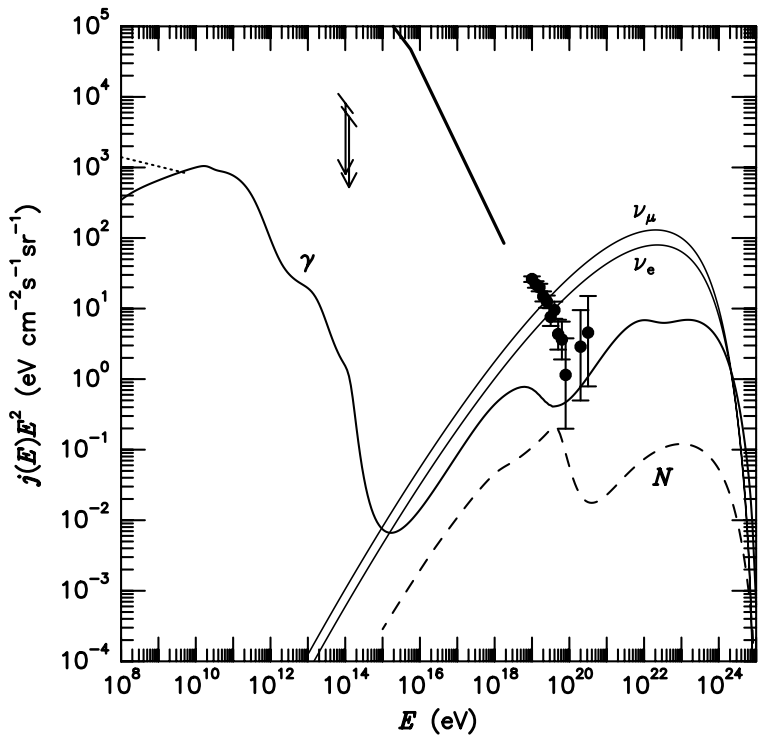,height=3.8in}
\medskip
\caption[...]{Predictions for the average differential fluxes
of $\gamma$-rays, nucleons and muon and electron neutrinos
by a typical TD scenario as described in the
text. We used the hadronic fragmentation function in modified
leading logarithmic approximation~\cite{detal}
for $m_X=2\times10^{25}\,$eV.
The average EGMF strength was assumed to be $10^{-12}\,$G.
Also shown are the combined data from the Fly's
Eye~\cite{Bird} and the AGASA~\cite{AGASA}
experiments above $10\,$EeV
(dots with error bars), piecewise power law fits to the observed
charged CR flux (thick
solid line) and experimental upper limits on the $\gamma$-ray
flux below $10\,$GeV from~\cite{EGRET}
(dotted line on left margin). The arrows indicate limits on the
$\gamma$-ray flux from.~\cite{Karle}
\label{F2}}
\end{figure}

Fig.~\ref{F2} shows the results for the $\gamma$-ray and nucleon
fluxes from a typical TD scenario, assuming an
EGMF of average strength $10^{-12}\,$G, along with
current observational constraints on
the $\gamma$-ray flux. The spectrum was normalized in the best
possible way to allow for an explanation of the observed HECR
events, assuming their consistency with a nucleon or
$\gamma$-ray primary (although a $\gamma$-ray primary is
somewhat disfavored~\cite{HVSV}). The flux below a few tens of
EeV is presumably caused by
conventional acceleration. The likelihood significance of the
fit (see~\cite{SLSB} for details) in
Fig.~\ref{F2} is $\simeq50\%$ for the energy range above
$100\,$EeV. While the shapes of
our spectra are similar to the ones obtained in,~\cite{PS}
this is in contrast to their procedure of normalizing to the
observed differential flux at $300\,$EeV which overproduces the
integral flux at higher energies.

Since the comparatively large amount of energy injected at high
redshifts is recycled to lower energy $\gamma$-rays, TD models
are significantly constrained~\cite{Chi,SJSB} by current
limits on the diffuse $\gamma$-ray background in the $100\,{\rm
MeV}-10\,{\rm GeV}$ region.~\cite{EGRET}
Note that the IR/O background strongly depletes the $\gamma$-ray
flux in the range $10^{11} - 10^{14}\,$eV, recycling it to
energies below $10\,$GeV (see Fig.~\ref{F2}).
Constraints from limits on CMB distortions and light
element abundances from $^4$He-photodisintegration are
comparable to the bound from the directly observed
$\gamma$-rays.~\cite{SJSB} The scenario in Fig.~\ref{F2}
obeys all current constraints within the normalization
ambiguities and is therefore quite viable.

Whereas the UHE nucleon and $\gamma$-ray fluxes are independent
of cosmological evolution, the $\gamma$-ray flux
below $\simeq 10^{11}\,$eV and the neutrino flux are proportional
to the total energy
injection which, for all other parameters held fixed, increases
with decreasing $p$.~\cite{SJSB} For $m_X=2\times10^{25}\,$eV,
scenarios with $p\la1$ are therefore ruled
out (see Fig.~\ref{F2}), whereas
constant comoving injection rates ($p=2$) are well within the
limits. Since the EM flux above $\simeq10^{22}\,$eV is
efficiently recycled to lower energies, this constraint turns
out to be basically independent of $m_X$ in case of a low
EGMF,~\cite{SLC1} in contrast to earlier analytical estimates
based on the continuous energy loss approximation
which underestimates the flux around $100\,$EeV.~\cite{Chi,SJSB}
The constraints from the flux limits below
$10\,$GeV become somewhat tighter for an EGMF of strength
$\ga10^{-11}\,$G.

The predicted neutrino fluxes~\cite{SLC2} are also consistent
with bounds from the Fr\'{e}jus experiment.~\cite{Rhode} At
these flux levels, neutrinos are unlikely candidates for the
observed HECR events due to their small interaction probability
in the atmosphere.~\cite{SL} A future detection of an appreciable
neutrino flux above $\sim1\,$EeV, for example, by a km$^3$ scale
neutrino observatory~\cite{GHS} could establish an experimental lower
limit on the ratio of energy injected as neutrinos versus
electromagnetically interacting particles and thus probe GUT
scale physics.~\cite{SLC2} In the scenario shown in Figs.~\ref{F1}
and~\ref{F2}, this ratio is about 0.3.

Our simulations show that
an isotropic $\gamma$-ray to total CR ($\gamma$/CR) flux ratio
at $\simeq10\,$EeV as high as $10\%$ can be attained.
However, this is only possible if a TD mechanism of the type
discussed above is responsible for most of the HECR above
$\simeq100\,$EeV, and if the EGMF is weaker than
$\simeq10^{-11}\,$G on scales of a few to tens of Mpc. In case
of acceleration sources of the HECRs, such high $\gamma$/CR flux
ratios at $\simeq10\,$EeV can only be attained in the direction
of powerful nearby acceleration sources. In contrast to the TD
models, in this case the $\gamma$-rays would be produced as
secondaries of nucleons interacting with the CMB. The spectral
shape of the $\gamma$-ray flux also depends on the EGMF which
determines where the energy loss of $\gamma$-rays is dominated
by synchrotron loss rather than inverse Compton scattering on
the CMB and thus could be used to ``measure'' the EGMF in the
range $\simeq10^{-10}-10^{-9}\,$G.~\cite{LOS}

Information on the EGMF structure could also be obtained by
observing the energy and and arrival time distribution of
nucleons from sources which release UHE cosmic rays on a time scale
short compared to $\simeq1\,$yr, i.e. in a
burst.~\cite{WME,SSLCH,LSOL}. This is because the average time delay
caused by deflection in the EGMF is
$\simeq0.9(E/100\,{\rm EeV})^{-2}(r/10\,{\rm
Mpc})^2(B/10^{-11}\,{\rm G})^2(l_c/1\,{\rm Mpc})\,{\rm yr}$,
where $E$ is the observed nucleon energy, $r$ is the source
distance, and $B$ and $l_c$ is the typical strength and the
coherence length of the EGMF, respectively. Some TD scenarios
such as the ones involving collapsing cosmic string
loops~\cite{BR} would imply the existence of such bursting
sources.

\section{Conclusions}
Some TD type scenarios of HECR origin are still
unconstrained by
current data and bounds on $\gamma$-ray and UHE CR fluxes.
For example, in case of an EGMF $\la10^{-9}\,$G, spatially
uniform annihilation of magnetic monopoles and antimonopoles is
still a viable model for GUT scales up to $10^{16}\,$GeV. In
such scenarios, the flux would be dominated by $\gamma$-rays
above $\simeq100\,$EeV and the possibility of a gap in the
spectrum arises.
A solid angle averaged $\gamma$/CR flux ratio at the $10\%$
level at $\simeq10\,$EeV is a signature of a
non-acceleration origin of HECRs hinting to the presence
of a TD mechanism. At the same time it would put an independent
new upper limit of $\simeq10^{-11}\,$G on
the poorly known EGMF on scales of a few to tens of Mpc.

The fact that some TD mechanisms would imply bursting sources could
provide another ``probe'' of the EGMF.

\section*{Acknowledgments}
I acknowledge my collaborators in the work on which this article
is based, namely, Venya Berezinsky, Pijush Bhattacharjee,
Paolo Coppi, Christopher Hill, Karsten Je\-dam\-zik, Sangjin Lee,
Martin Lemoine, Angela Olinto, and David Schramm.
This work was supported by the DoE, NSF and NASA at the
University of Chicago, by the DoE and by NASA through grant
NAG 5-2788 at Fermilab, and by the Alexander-von-Humboldt
Foundation.


\section*{References}

\begin {thebibliography}{900}

\bibitem{Bird} D.~J.~Bird {\it et al.}, {\em Phys.~Rev.~Lett.}
{\bf 71}, 3401 (1993);
{\em Astrophys.~J.} {\bf 424}, 491 (1994); {\em ibid.} {\bf
441}, 144 (1995).

\bibitem{AGASA} N.~Hayashida {\it et al.}, {\em
Phys.~Rev.~Lett.} {\bf 73}, 3491
(1994); S.~Yoshida {\it et al.}, {\em Astropart.~Phys.} {\bf 3},
105 (1995).

\bibitem{Hillas} A.~M.~Hillas, {\em
Ann. Rev. Astron. Astrophys.} {\bf 22}, 425 (1984).

\bibitem{GZK} K.~Greisen, {\em Phys.~Rev.~Lett.} {\bf 16}, 748 (1966);
G.~T.~Zatsepin and V.~A.~Kuzmin, {\em Pisma
Zh.~Eksp.~Teor.~Fiz.} {\bf 4}, 114
(1966) [{\em JETP.~Lett.} {\bf 4}, 78 (1966)].

\bibitem{Puget} J.~L.~Puget, F.~W.~Stecker, and J.~H.~Bredekamp,
{\em Astrophys. J.} {\bf 205}, 638 (1976).

\bibitem{SSB} G.~Sigl, D.~N.~Schramm, and P.~Bhattacharjee,
{\em Astropart. Phys.} {\bf 2}, 401 (1994).

\bibitem{ES} J.~W.~Elbert and P.~Sommers, {\em Astrophys. J.}
{\bf 441}, 151 (1995).

\bibitem{BHS} P.~Bhattarcharjee, C.~T.~Hill, and D.~N.~Schramm,
{\em Phys.~Rev.~Lett.} {\bf 69}, 567 (1992).

\bibitem{detal} Yu.~L.~Dokshitzer, V.~A.~Khoze, A.~H.~M\"uller,
and S.~I.~Troyan, {\em Basics of
Perturbative QCD} (Editions Frontieres, Singapore, 1991).

\bibitem{Hill} C.~T.~Hill, {\em Nucl. Phys.} B {\bf 224}, 469 (1983).

\bibitem{BR} P.~Bhattacharjee and N.~C.~Rana, {\em Phys.~Lett.}
B {\bf 246}, 365 (1990).

\bibitem{BS} P.~Bhattarcharjee and G.~Sigl, {\em Phys.~Rev.} D
{\bf 51}, 4079 (1995).

\bibitem{GK} A.~J.~Gill and T.~W.~B.~Kibble, {\em Phys.~Rev.} D
{\bf 50}, 3660 (1994).

\bibitem{Lee} S.~Lee, report FERMILAB-Pub-96/066-A,
astro-ph/9604098, submitted to {\em Phys.~Rev.} D.

\bibitem{ABS} F.~A.~Aharonian, P.~Bhattacharjee, and D.~N.~Schramm,
{\em Phys.~Rev.} D {\bf 46}, 4188 (1992).

\bibitem{SLSB} G.~Sigl, S.~Lee, D.~N.~Schramm, and
P.~Bhattacharjee, {\em Science} {\bf 270}, 1977 (1995).

\bibitem{Bird1} D.~J.~Bird {\it et al.}, Proceedings of the 24th
International Cosmic-Ray Conference (Rome, 1995), vol.~2,
p.~504.

\bibitem{Proc} M.~Boratav {\it et al.}, Eds., Proceedings of the
{\em International Workshop on Techniques to Study Cosmic Rays
with Energies} $\geq10^{19}\,$eV, {\em Nucl. Phys.} B
(Proc. Suppl.) {\bf 28B} (1992).

\bibitem{Chi} X.~Chi {\it et al.}, {\em Astropart. Phys.} {\bf 1}, 129
(1993); {\em ibid.} {\bf 1}, 239 (1993).

\bibitem{SJSB} G.~Sigl, K.~Jedamzik, D.~N.~Schramm, and V.~Berezinsky,
{\em Phys.~Rev.} D {\bf 52}, 6682 (1995).

\bibitem{PS} R.~J.~Protheroe and T.~Stanev, report ADP-AT-96-6,
astro-ph/9605036, submitted to {\em Phys. Rev. Lett.}

\bibitem{SLC1} G.~Sigl, S.~Lee, and P.~Coppi, report
FERMILAB-Pub-96/087-A, astro-ph/9604093, submitted to
{\em Phys.~Rev.~Lett.}

\bibitem{HVSV} F.~Halzen, R.~A.~Vazquez, T.~Stanev, and
H.~P.~Vankov, {\em Astropart. Phys.} {\bf 3}, 151 (1995).

\bibitem{EGRET} A.~Chen, J.~Dwyer, and P.~Kaaret, {\em
Astrophys. J.} {\bf 463}, 169 (1996); C.~E.~Fichtel, Proceedings
of the 3rd Compton Observatory Symposium, {\em
Astron. Astrophys. Suppl.}, in press (1996).

\bibitem{SLC2} G.~Sigl, S.~Lee, D.~N.~Schramm, and P.~Coppi,
report astro-ph/9610221, submitted to {\em Phys.~Lett.~B}.

\bibitem{Rhode} W.~Rhode {\it et al.}, {\em Astropart. Phys.}
{\bf 4}, 217 (1996).

\bibitem{SL} G.~Sigl and S.~Lee, Proceedings of the 24th
International Cosmic-Ray Conference (Rome, 1995), vol.~3,
p.~356.

\bibitem{GHS} see, e.g., T.~K.~Gaisser, F.~Halzen, and T.~Stanev,
{\em Phys. Rep.} {\bf 258}, 173 (1995).

\bibitem{LOS} S.~Lee, A.~V.~Olinto, and G.~Sigl, {\em
Astrophys. J.} {\bf 455}, L21 (1995).

\bibitem{WME} E.~Waxman and J.~Miralda-Escud\'{e}, preprint
astro-ph/9607059, submitted to Astrophys. J. Lett.

\bibitem{SSLCH} G.~Sigl, D.~N.~Schramm, S.~Lee, P.~Coppi, and
C.~T.~Hill, preprint FERMILAB-Pub-96/121-A, astro-ph/9605158.

\bibitem{LSOL} M.~Lemoine, G.~Sigl, A.~V.~Olinto, and S.~Lee,
in preparation.

\bibitem{Karle} A.~Karle {\it et al.}, {\em Phys.~Lett.} B {\bf
347}, 161 (1995).

\end{thebibliography}
\end{document}